# Improving overlay maps of science: combining overview and detail

Peter Sjögårde[a,b,x]

[a]Health Informatics Centre, Department of Learning, Informatics, Management and Ethics, Karolinska Institutet, Stockholm, Sweden
[b]University Library, Karolinska Institutet, Stockholm, Sweden

ORCID:
[x]https://orcid.org/0000-0003-4442-1360

Email: peter.sjogarde@ki.se

Corresponding author: Peter Sjögårde, University Library, Karolinska Institutet, 17177 Stockholm, Sweden

## Abstract

Overlay maps of science are global base maps over which subsets of publications can be projected. Such maps can be used to monitor, explore, and study research through its publication output. Most maps of science, including overlay maps, are flat in the sense that they visualize research fields at one single level. Such maps generally fail to provide both overview and detail about the research being analyzed. The aim of this study is to improve overlay maps of science to provide both features in a single visualization.

I created a map based on a hierarchical classification of publications, including broad disciplines for overview and more granular levels to incorporate detailed information. The classification was obtained by clustering articles in a citation network of about 17 million publication records in PubMed from 1995 onwards.

The map emphasizes the hierarchical structure of the classification by visualizing both disciplines and the underlying specialties. To show how the visualization methodology can help getting both overview of research and detailed information about its topical structure, I projected two overlay maps onto the base map: (1) open access publishing and (2) coronavirus/Covid-19 research.

# 1 Introduction

To be able to support and manage research activities there is a need to monitor and study research, for example to coordinate research, follow-up investment or strengthen collaboration in targeted areas. It is relatively easy to keep track of the research activities of small research units. In contrast, large research units, for example whole universities, may have hundreds or thousands of employees, producing thousands of research publications each year within a broad spectrum of topics. Keeping track of the competences, research areas and collaborations is a challenging task for such organizations.

Research publications are one important output of research activity. Publications can be studied to monitor research activity and gain insight into aspects such as collaboration patterns, specialization, strong research areas, trends, and development. Overlay maps of science have been proposed to "offer an intuitive way of visualizing the position of organizations or topics in a fixed map" (Rafols et al., 2010). Overlay maps are base maps over which subsets of publications or filters can be projected, for example to study the position of organizations or topics, or to highlight properties such as citation impact, open access publishing or clinical research (Kay et al., 2014).

To this point, most overlay maps have been flat in the sense that they visualize research fields at one single level, commonly the levels of research disciplines or specialties. Such maps generally fail to provide *both* overview and detail about the research being studied. The aim of this study is to improve overlay maps of science to provide these two features in one single, interactive map, having a focus on the biomedical sciences. The maps created enable users to explore multiple levels of a hierarchical classification in a single interactive visualization.

## 2 Background

Visualizations of science have been around for a long time (for overviews, see Börner et al., 2005; Petrovich, 2020; van Eck & Waltman, 2014; Zitt et al., 2019). Early work focused on maps restricted to one or a few research domains. Different aspects of domains have been visualized and studied using a variety of entities and relations, for example: co-publishing between researchers or organizations, co-occurrence of keywords, and citation relations between publications or journals.

Beginning in the end of the 1990s, maps that cover large parts of the science system (at least within the natural science and biomedicine) have been created. Initial maps of science were based on journals and made it possible to get unpreceded overviews of the science system (for examples, see Bassecoulard & Zitt, 1999; Boyack et al., 2005; Leydesdorff, 2004, 2006; Moya-Anegón et al., 2004).

Rafols et al. (2010) showed how comprehensive maps of science can be used as base maps, over which overlays can be projected. The idea of such a map is to fixate the positions of the nodes in the map, representing for example research fields, so that an overlay projected onto the map can be easily compared with the base map, as well as with other projections. For instance, consider the publication outputs, A and B, of two universities. An overlay map is created by projecting A onto a base map. The size of the nodes is scaled in relation to the distribution of A over research fields. Another map is created based on B, using the same procedure. We can now compare the subject orientation of the two universities by exploring the two maps and compare node sizes. If we color the nodes of the maps based on some variable, we can analyze different aspects of A and B, for example the amount of open access publishing, citation impact or degree of international collaboration in different research fields. Compared to maps restricted to particular domains, overlay maps provide context and points of reference, for example by offering the possibility to spot areas in which A and B do not have any research.

Since overlay maps were introduced in scientometrics, they have been used in many applications; Kay et al. (2014) used overlay maps to visualize patents by companies, Tank and Shapira (2011) analyzed the growth of US and China co-publications in nanotechnology, Klaine et al. (2017) positioned environmental, health, and safety of nanomaterials in relation to general nanotechnology, Leydesdorff et al. (2015) created an overlay map based on journal relations using Scopus data and exemplified how the map can be used to explore the publication output of authors, organizational units or other publication sets, and Rotolo et al. (2017) used three case studies to demonstrate the use of overlay maps for strategic intelligence.

Most of the applications have used maps at one single level and one single entity (e.g., journals or keywords). The study by Rotolo et al. (2017) includes what they refer to as "cognitive" base maps of research publications at different granularity levels. These maps are based on the Web of Science categories at the broadest level, journals at the meso-level and medical subject headings (MeSH) at the most granular level. However, the different granularity levels are presented in different maps and hierarchical relations between levels are not shown.

To provide the possibility to navigate from broad to narrow levels in one single map, I base the map presented in this paper on a hierarchical classification obtained by clustering articles in a citation network.

Publication-level classification at a global level (covering a complete multidisciplinary data source) obtained by clustering articles in a citation network was first implemented by Waltman and van Eck (2012). Compared to classification at the journal-level, publication-level classifications can be made more granular. A hierarchical structure can be obtained by merging clusters at lower levels into broader clusters. Publication-level classifications have been used to create overlay maps (RoRI Institute, 2019). However, the applications are few and lack hierarchical structure, other than node coloring by major research areas.

The validity of the clustering solutions created by clustering articles in citation networks has been contested (Held et al., 2021). There is no ground truth classification and different methodological choices results in different, sometimes equally valid, representations of research delineation (Glänzel & Schubert, 2003; Gläser et al., 2017; Klavans & Boyack, 2017; Mai, 2011; Sjögårde & Ahlgren, 2018; Smiraglia & van den Heuvel, 2013; Velden et al., 2017; Waltman & van Eck, 2012). Nevertheless, the results have been compared to a wide range of baselines and many different applications have been evaluated and compared (Ahlgren et al., 2020; Boyack, 2017; Boyack et al., 2011; Boyack & Klavans, 2010, 2010, 2018; Donner, 2021; Haunschild et al., 2018; Sjögårde & Ahlgren, 2018, 2020; Šubelj et al., 2016; Waltman et al., 2020). Overall, the findings of this research literature suggest that publication-level classifications based on citation relations perform at least reasonably well when evaluated quantitatively, but there is still a lack of qualitative evaluations.

Including citations that are external to the analyzed data set improves accuracy of a clustering solution (Ahlgren et al., 2020; Boyack, 2017; Donner, 2021; Klavans & Boyack, 2011). This is an advantage of a global approach, compared to a local one in which a clustering is based on a restricted set of publications. Nonetheless, a local approach may be preferable in some applications because it can emphasize the local, within field, context of research. However, local maps are difficult to compare to other maps. The rationale of overlay maps is to make comparisons between maps. A global approach is therefore most often a superior approach for overlay maps.

# 3 Data and method

To create a visualization of biomedical research literature that incorporates both overview and detail I based the visualization on a hierarchical publication-level classification. This classification was obtained by clustering articles in a (direct) citation network of PubMed records. Currently, PubMed indexes over 1 million publications yearly, covering a wide range of biomedical research disciplines. I therefore refer to the classification created as “global”, even though it does not have a comprehensive coverage of all fields of science.

A similar classification was recently created by Boyack et al. (2020). This classification differs mainly by the choice of similarity measure between publications. Boyack et. al based their classification on a combination of direct citations and textual similarity. By complementing direct citations with textual similarity, they were able to include publications that would otherwise have no relations. Thus, the relation between publications in their approach is a mixture of different, fundamentally different, similarity measures and makes the interpretation of the classification more difficult. Since the model of Boyack et al. was published, more citations have been made openly available and it is now possible to create citation-based classifications with a more comprehensive coverage. I therefore base my classification strictly on direct citations.

Figure 1 illustrates the process used to obtain the classification from the citation network and to create a base map from the classification. The base map incorporates features to emphasize the hierarchical structure of the classification. In the next section (3.1), I describe the data and process to obtain the classification, and in the following section (3.2), I describe the process to create the base map from the classification.

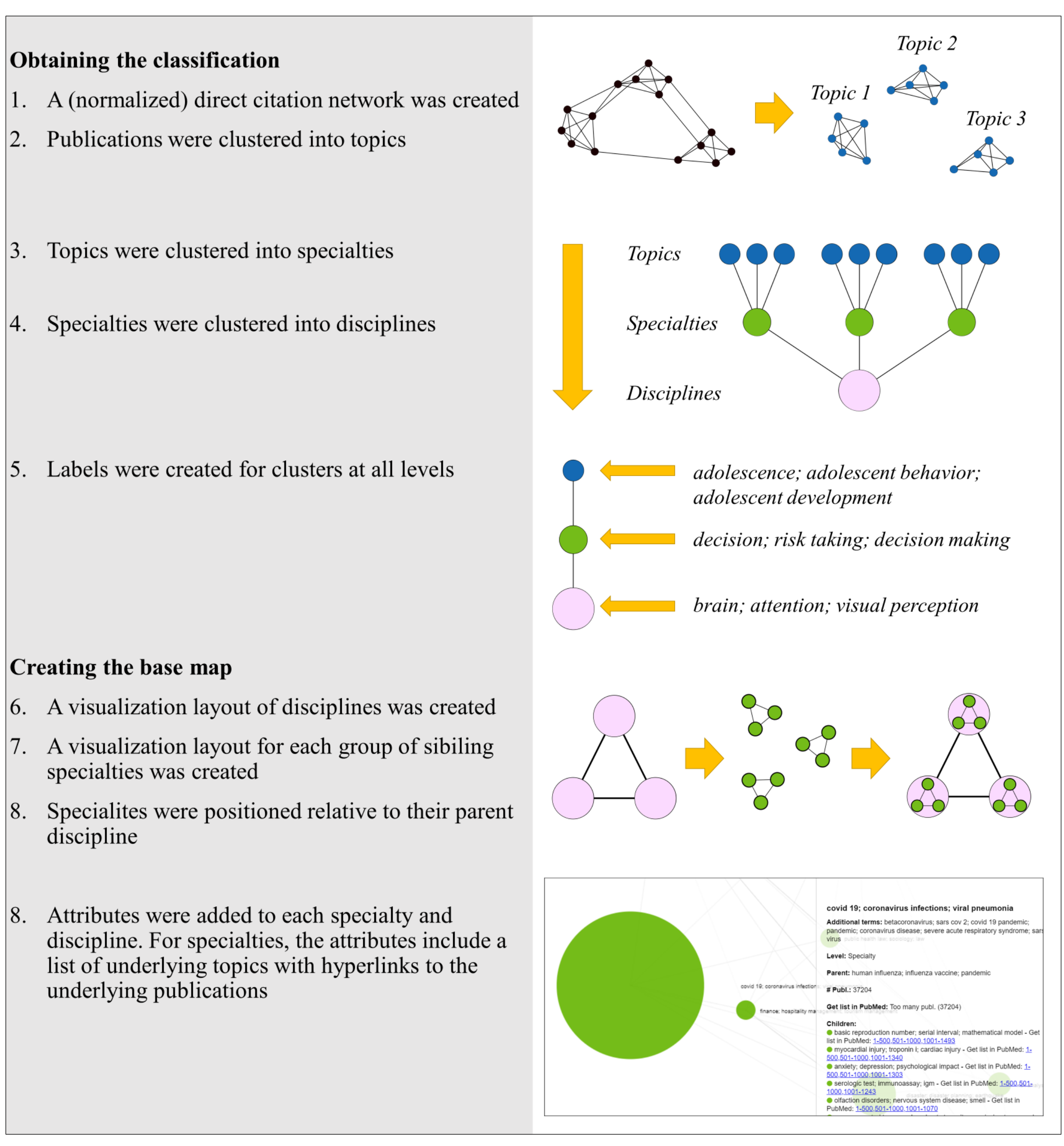

*Figure 1: Illustration of the process used to create and visualize the classification.*

## 3.1. Obtaining the classification

I used PubMed data to create a classification of publications in four levels based on citation relations from the NIH Open Citation Collection (Hutchins et al., 2019; iCite, 2019). I used the bibliometric system at Karolinska institutet for the analysis. The system contains PubMed data from 1995 onwards. Data were extracted in February 2021 (version 14 of the NIH Open Citation

Collection) and restricted the data to the publication types "article" and "review", a total of about 17 million publications with about 380 million direct citation relations. In the remainder of this paper, I use the term "publication" to refer to both articles and reviews.

Except for some modifications, I obtained the classification using the methodology put forward in Waltman and van Eck (2012). In accordance with this methodology, direct citation relations were used to create a network. Citation relations were normalized in relation to each publication's total number of citation relations. The Leiden algorithm (Traag et al., 2019) was used to obtain a partitioning of the publications.[1] To get clusters of substantial size, I restricted the cluster size to a minimum of 50 publications by reclassifying publications in clusters below this minimum size using the method provided in the software. The resolution parameter was calibrated to obtain clusters of about the same size as in Sjögårde & Ahlgren (2018) for the corresponding publication years, resulting in 60,650 clusters. Thereby, clusters approximately correspond to research *topics,* and I refer to clusters at this level as such.

Topics were clustered into larger groups based on their relatedness using the methodology in Sjögårde and Ahlgren (2020) and calibrating the resolution parameter to obtain clusters of about the same size for corresponding publication years as obtained in the referred article. Thereby, clusters approximately correspond to research *specialties*. A minimum threshold of at least 500 publications was used at this level, which resulted in 2,139 specialties.[2]

The next level approximates the level of research *disciplines*. My intention was to create clusters of approximately the size of other broad classifications, such as Web of Science journal categories (about 250 clusters), Science Metrix journal classification (180 clusters at the "subfields" level) and Scopus Subject areas (about 330 clusters). Because the classification only includes biomedicine I aimed for a smaller number of clusters. I tested several different values of the resolution parameter. Too low values merged specialties with seemingly weak relatedness into coarse clusters, while too high values resulted in many specialties being unmerged. I finally chose a solution with 169 disciplines, after restricting cluster sizes to at least 100,000 publications.

---

[1] The process took about 8h 25min to run 100 iterations. 256 GB RAM was allocated for the process. The value of the quality function (CPM) was 0.427. The resolution parameter was set to 0.00010. Version 1.1.0 of the software was downloaded from https://github.com/CWTSLeiden/networkanalysis [2020-11-20].

[2] At aggregated level reclassification was performed by merging clusters below the threshold with the cluster above the threshold having the strongest relational strength.

Disciplines were grouped into 23 broader research areas. It is particularly difficult to obtain good labels at this level, because terms extracted from bibliographic fields tend to be too narrow. For this reason, the research areas are not displayed in the visualization. The research areas are used in the visualization only to keep sibling disciplines (the group of disciplines with the same parent) in proximity.

To create labels, I used the procedure proposed in Sjögårde et al. (2021). Noun phrases were extracted from article titles, medical subject headings (MeSH), journal titles and author addresses. A noun phrase was operationalized as a sequence of adjectives and nouns, ending with a noun (van Eck et al., 2010). A Java-program was written for this purpose and the Stanford Core NLP software was used for data mining (Manning et al., 2014), in particular the lemmatizer and the Part-Of-Speech tagger (Toutanova et al., 2003; Toutanova & Manning, 2000).[3] The relevance of terms to clusters was calculated using term frequency to specificity ratio (TFS, Sjögårde et al., 2021). TFS balances term frequency and term specificity to obtain terms that are both frequent in a cluster and specific to the cluster. For each cluster, the three terms with the highest TFS value were concatenated into a label. Seven more terms are listed when clicking on a node in the visualization. I used article titles and MeSH to create labels at the topic level (α=0.33 was used for the TFS calculation). I used article titles, MeSH and journal titles at the specialty level (α=0.5) and journal titles and author addresses at the discipline level (α=0.67).

[3] Stanford CoreNLP is available at http://stanfordnlp.github.io/CoreNLP/

## 3.2. Creating the base map

As a basis for the map, I created a network of disciplines. In the following step I created a network of underlying specialties for each discipline. This network was positioned to the discipline node. A list of topics was created for each specialty. This list is displayed when clicking a node. In the following I describe the steps in detail:

*A. Discipline level*

A1. I created a list of disciplines with the attributes shown in Table 1. For the purpose of illustration, the table presents values for an example discipline. The size attribute was calculated as the square root of the number of publications. This makes the area of each node proportional to the number of publications. If a cluster contained no more than 500 publications, a hyperlink was created to the underlying publications ("Get list in PubMed"). If a cluster contained 501-5,000 publications a separate hyperlink was created for each batch of 500 publications (1-500, 501-1,000, etc). If more than 5,000 publications no hyperlinks were provided and instead the text "Too many publ." was shown. This was done because of restrictions in hyperlink length. The underlying specialties were listed in the column "Children". Labels and numbers of publications for the specialties were concatenated into a list. The interactive map contains hyperlinks to the underlying publications for each of the specialties (with the same restriction to 5,000 publications). Discipline nodes were colored according to their cluster at the top level (research areas).

*Table 1: Attributes for an example discipline. Hyperlinks omitted.*

| Attribute | Value |
|---|---|
| **id** | l3.14 |
| **label** | dermatology; melanoma; skin |
| **size** | 178.2 |
| **color** | rgba(104,14,75,0.5) |
| **Additional terms** | psoriasis; pathology; atopic dermatitis; skin disease; venereology; keratinocyte; disease |
| **Level** | Discipline |
| **# Publ.** | 31763 |
| **Get list in PubMed** | Too many publ. (31763) |
| **Children** | ● skin neoplasm; pigmented nevus; malignant melanoma - # Publ.: 3825<br>● psoriasis; psoriatic arthritis; rheumatology - # Publ.: 3762<br>● atopic dermatitis; pruritus; allergy - # Publ.: 2981<br>● healing; wound healing; keloid - # Publ.: 2887<br>● alopecia; alopecia areata; hair - # Publ.: 1859<br>● melanin; monophenol monooxygenase; melanocortin - # Publ.: 1815<br>[…] |
| **x** | 561.2 |
| **y** | 551.5 |

A2. The ForceAtlas algorithm (Jacomy et al., 2014) was used to create a layout based on the average normalized direct citation value between the nodes (the same relatedness value used for clustering). [4] The algorithm resembles a physical system in which nodes repulse each other and edges attract the nodes. The magnitude of the attraction is relative to the weight of the edges. To emphasize the hierarchical structure, relations between sibling disciplines were scaled by a factor set to 3. Thereby, siblings were positioned closer to each other, which facilitated the interpretation of the network. Finally, x- and y-coordinates were supplied to the list of disciplines.

[4] An R-function (with base code in C) was created by my colleague Robert Juhasz for this task. The function is equivalent to the Force Atlas layout in Gephi . The following parameter values were used: number of iterations = 10,000, inertia = 0.1, repulsion strength = 500, attraction strength = 10, max displacement = 10, freeze balance = true, freeze strength = 80, freeze inertia = 0.2, gravity = 30, outbound attraction distribution = false, adjust sizes = false, speed = 1, cooling = 1.

*B. Specialty level*

B1. For each discipline I created a network of its children, with each child representing one of the underlying specialties of the disciplines. Edges between specialties were assigned a weight based on the average normalized direct citation value between specialties. The ForceAtlas algorithm was used to create a layout for each set of sibling specialties.[5]

B2. Each network of sibling specialties was positioned in such a way that its midpoint was placed at the coordinates of the parent discipline, and its expansion was scaled to approximately fit the size of the parent node. The midpoint was defined as the average of the x values and y values respectively. Formally, the specialties were positioned as follows. Let $\mathbf{V} = \{v_1, ..., v_n\}$ be a set of sibling specialties, with the parent $p$. $x(v_i)$, $y(v_i)$ are the initial coordinates of $v_i$ and $x(p)$, $y(p)$ are the coordinates of $p$. The adjusted x-coordinates of $v_i$ are given by:

$$x_a(v_i) = \left( x(v_i) - \frac{\sum_{i \in \{1,...,n\}} x(v_i)}{n} \right) \times m \times (1/n) + x(p)$$

where *m* is a constant parameter used to scale the expansion of the network. *m* and *n* are used to scale the size of the network to approximately fit the size of the parent node. In a first attempt I tried to scale the expansion of the network in relation to the size of the parent node *p*. However, using the number of specialties *n* turned out to yield better results. *m* was set to 0.5 after some testing. The calculation of the y-coordinates of $v_i$ was performed analogously to the calculation of the x-coordinates.

---

[5] The following parameter values were used: number of iterations = 10,000, inertia = 0.1, repulsion strength = 2000, attraction strength = 20, max displacement = 10, freeze balance = true, freeze strength = 80, freeze inertia = 0.2, gravity = 10, outbound attraction distribution = false, adjust sizes = false, speed = 1, cooling = 1.

B3. The same attributes were calculated for specialties as for disciplines. Underlying topics are listed in the attribute "Children" in the case of specialties.

B4. Sizes of the specialty nodes were rescaled by dividing with 2. This was done for better readability of the visualization.

I visualized the networks using the sigma.js package created with the "SigmaExporter" plugin for the visualization software Gephi.[6] R was used to create networks and other files necessary for the visualization (a json network file,[7] a json configuration file[8] and an html-file).

The file size of the full base map is large due to the high amount of data in hyperlinks. To decrease loading time, I restricted the available online version of the map to the publication years 2018-2020, showing about 3.3 million publications. Below I refer to this map as the base map. As a result of this restriction, the map shows the current (or most recent) state of the biomedical literature.

# 4 Results

In this section I demonstrate how the base map provides both overview and detail by visualizing the hierarchical structure of the classification which it has been built upon. I then present two cases that show how the map can be used to enrich the study of research activities:

---

[6] The SigmaExporter was developed through the InteractiveVis project at the Oxford Internet Institute, University of Oxford. The Java code of the exporter is available under a GPLv3 License. https://gephi.org/plugins/#/plugin/sigmaexporter [2020-03-05]

[7] https://petersjogarde.github.io/pm_classification/2021/research_areas/data.json

[8] https://petersjogarde.github.io/pm_classification/2021/research_areas/config.json

(1) monitoring open access publishing in the biomedical literature over time and (2) studying coronavirus/Covid-19 research and its historical roots.

## 4.1. The base map

The base map shows distinctly colored groups of sibling disciplines. Figure 2 is a screen shot of the interactive map that is available online.[9] The largest groups revealed by the map are: cell and molecular biology, pharmacology and drug treatment, physiology, psychiatry, neuroscience and neurology, cardiovascular diseases, health professions, surgery, cancer diagnosis and treatment, infectious diseases, ecosystems and environmental medicine, biophysics and biochemistry, and immunology.

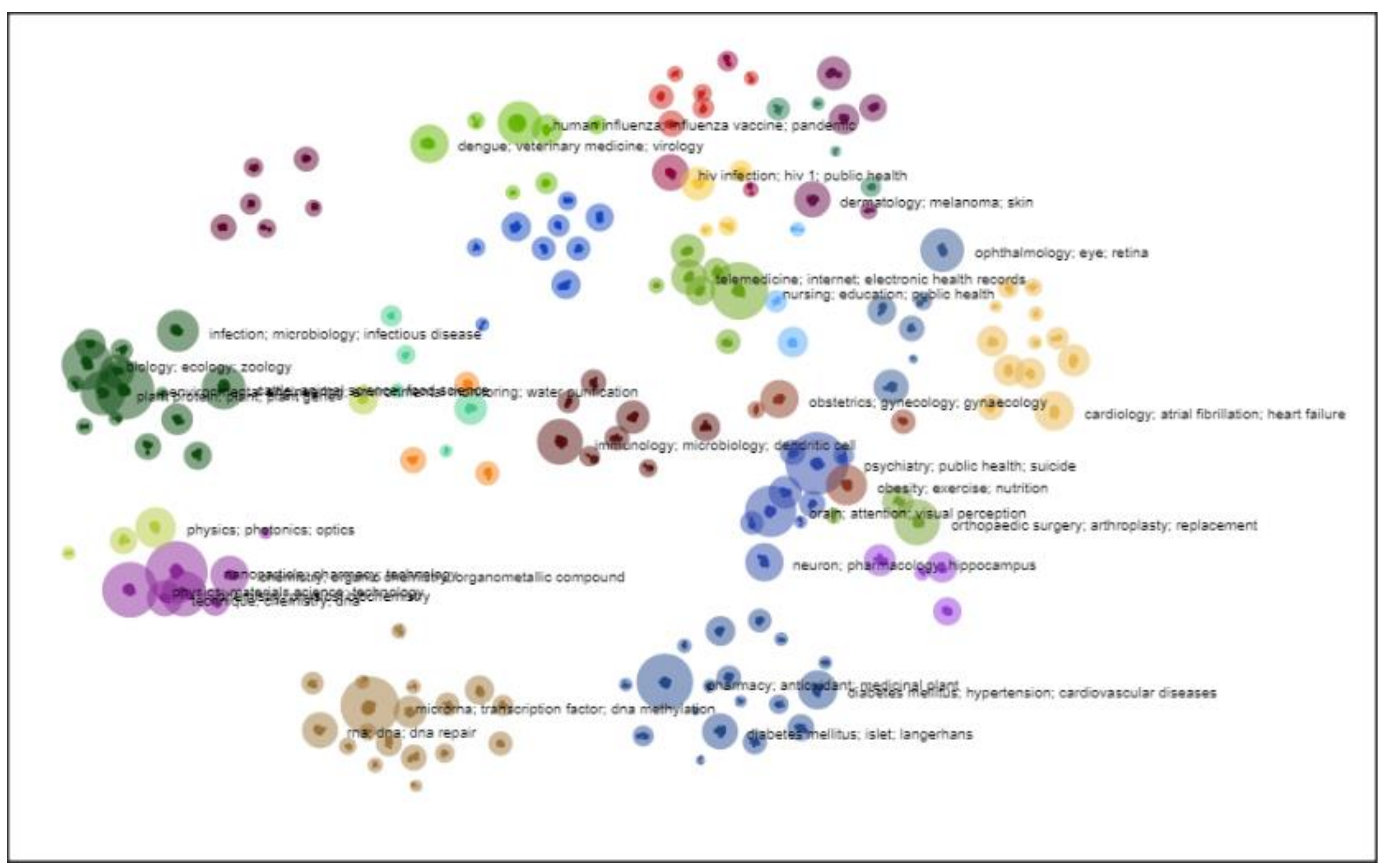

*Figure 2: A base map of biomedical science based on about 3.3 million articles from 2018-2020.*

The zooming feature is displayed in Figure 3, showing specialties in ”dermatology; melanoma; skin”. The figure reveals major specialties addressing skin cancer, psoriasis, allergy, skin treatment and hair loss. By looking at the node sizes we can estimate the relative size of these fields. Skin cancer is the largest node, but almost the same size as psoriasis. Some nodes are about half the size of these large nodes, for example hair loss (“alopecia; alopecia areata; hair”) and laser therapy (“laser therapy; laser; melasma”), while others are very small, for example nail conditions (“ingrown nails; yellow nail syndrome; ingrown toenail”) and skin diseases related to HIV (“hiv infection; eosinophilia; eosinophilic pustular folliculitis”).

[9] https://petersjogarde.github.io/pm_classification/2021/research_areas/index.html

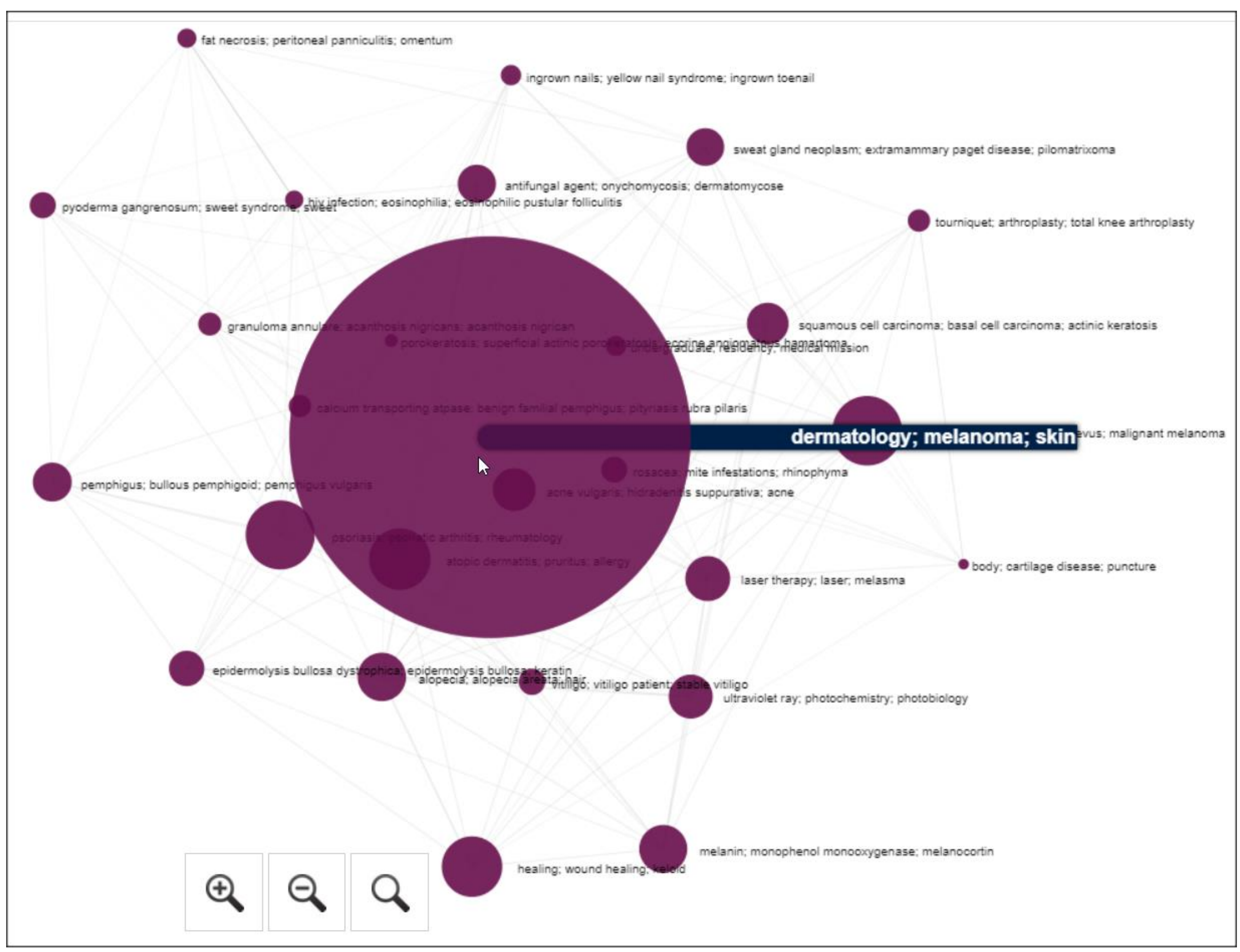

*Figure 3: The discipline ”dermatology; melanoma; skin” and its child nodes.*

Clicking on a specialty gives the user further information. This feature is exemplified in Figure 4, in which the information panel for the skin cancer specialty (“skin neoplasm; pigmented nevus; malignant melanoma”) is displayed. The information panel reveals sub-topics addressing treatment, behavior and risk factors, sentinel lymph nodes and different kinds of skin cancer, such as spitzoid melanoma and subungual melanoma. Hyperlinks make it possible to retrieve the publications underlying each topic in PubMed.

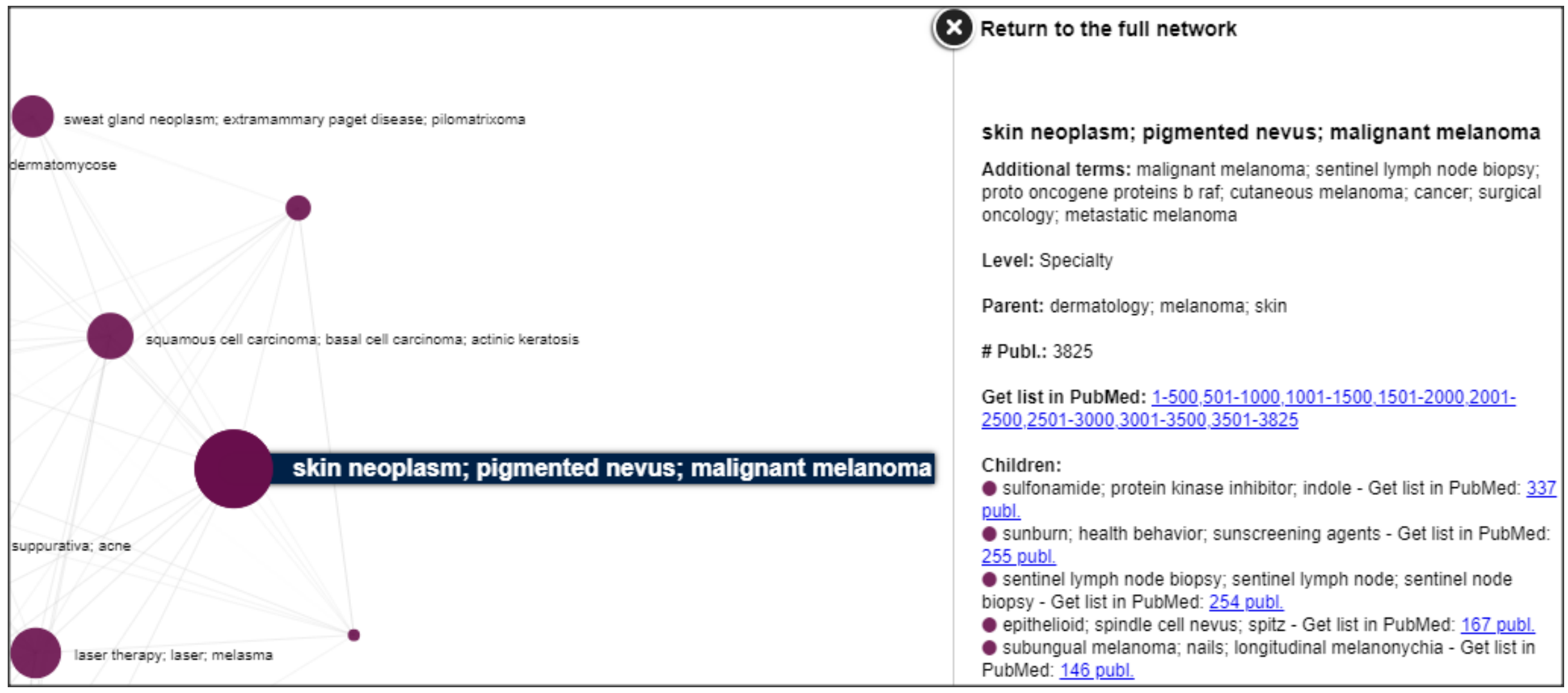

*Figure 4: The specialty "skin neoplasm; pigmented nevus; malignant melanoma". The information panel shows information about the specialty. The image has been cut to show the five largest topics in the specialty, the interactive map shows all topics.*

## 4.2. Open access

To show open access development over time three maps covering three time periods were projected onto the base map: 2008-2010 (Figure 5)[10], 2013-2015 (Figure 6)[11] and 2018-2020 (Figure 7)[12]. Nodes were colored by the level of articles openly available, ranging from 0% to 100%. Unpaywall was used to calculate the percentage of open access publications in each cluster.[13] A publication was regarded as open access if it had the status "gold", "bronze", "green" or "hybrid".

In the first time period (2008-2010) the average open access level in disciplines ranged from about 20% to 60%. High levels are seen in cell and molecular biology, for example the discipline “microrna; transcription factor; dna methylation” with 61.3% open access articles. Individual specialties had even higher rates, for example “chromatin immunoprecipitation; dna sequence analysis; bioinformatics” (77.1%) and “cell polarity; protein; insect protein” (77.3%). Other areas show low open access rates, for example in biochemistry where “nanoparticle; pharmacy; technology” had a rate of 23.4%, with some specialties ranging from 10% to 20%. In more applied areas the open access rates generally ranged from about 30% to 40%, for example in “cardiology; atrial fibrillation; heart failure” (37.5%) and “nursing; education; public health” (30.0%).

[10] https://petersjogarde.github.io/papers/hiervis/oa/2008-2010/index.html
[11] https://petersjogarde.github.io/papers/hiervis/oa/2013-2015/index.html
[12] https://petersjogarde.github.io/papers/hiervis/oa/2018-2020/index.html
[13] https://unpaywall.org/

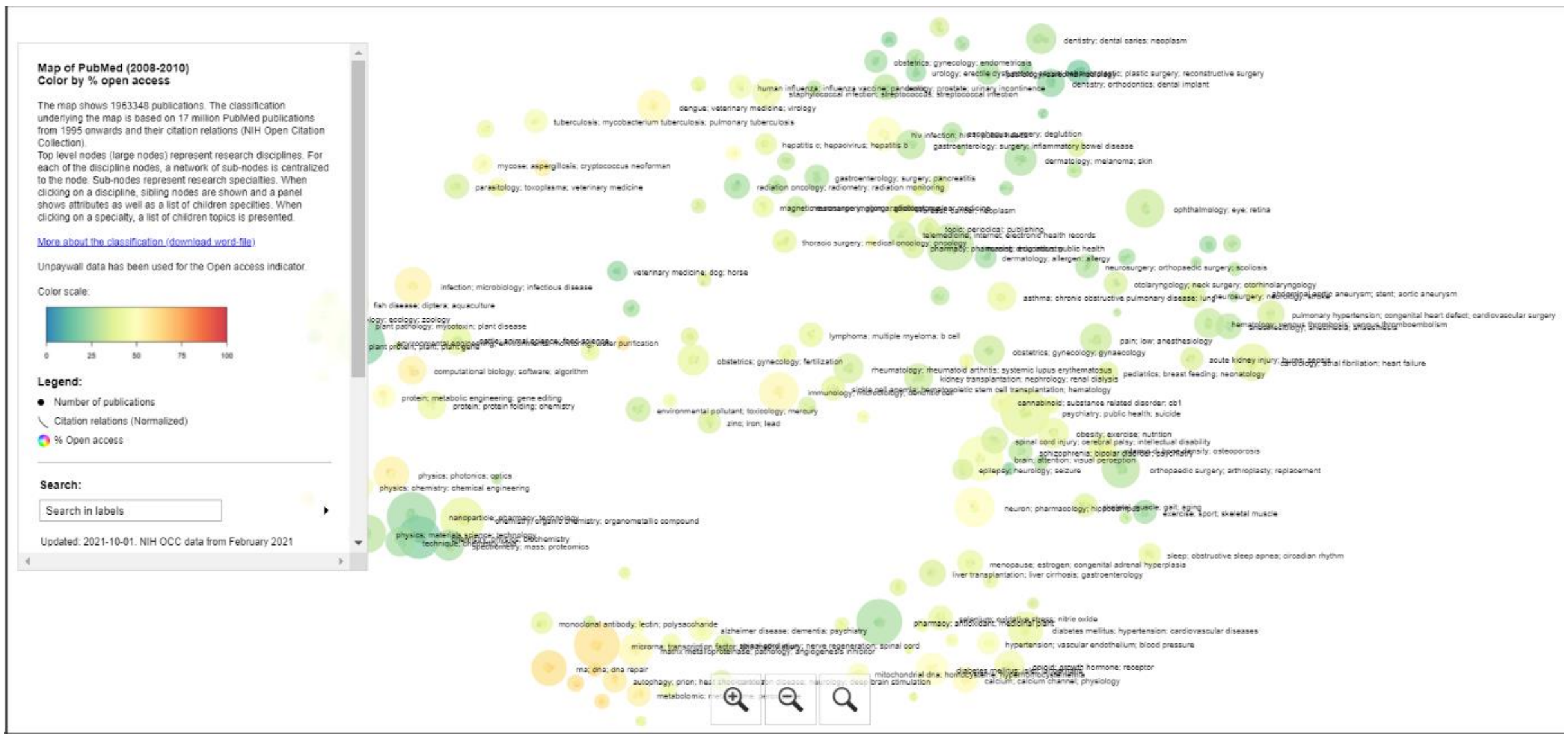

*Figure 5: Open access rates 2008-2010.*

Comparing 2008-2010 to 2013-2015 reveals a significant increase of the open access rates. Many areas had increased the open access rate by more than ten percentage points. The number of green nodes, signifying rates around 30%, was small in 2013-2015 and most were centralized to biochemistry (plastic surgery being an exception in the upper right corner).

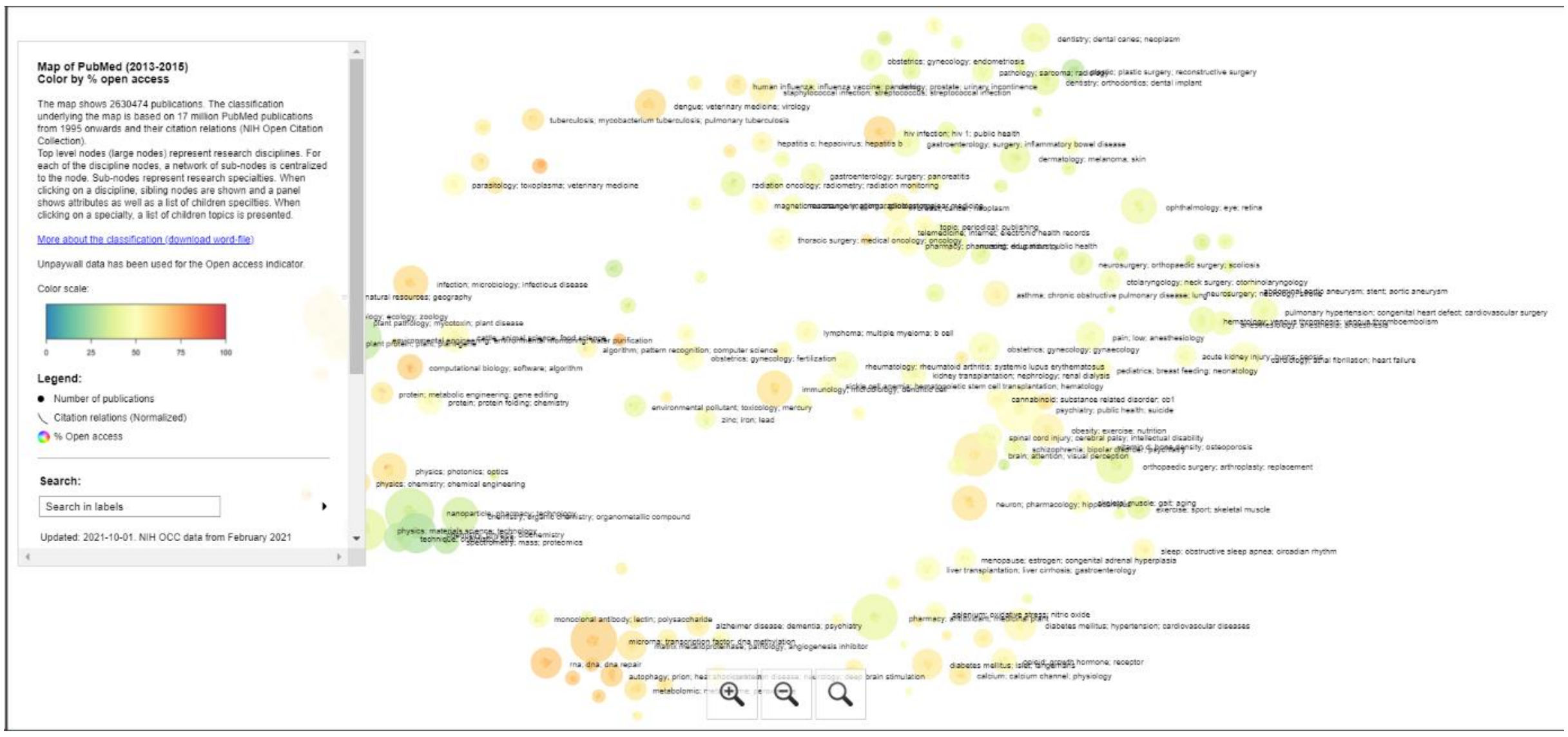

*Figure 6: Open access rates 2013-2015.*

The last time period (2018-2020) shows similar open access rates as the second time period. However, the group of biochemical disciplines got more yellow, signifying an increase of open access rates. At the same time the disciplines that had high rates in 2013-2015 did not increase further. The map for 2018-2020 shows a more equal distribution of open access rates over disciplines than the maps for the earlier time periods. The growth of coronavirus research due to the Covid-19 pandemic is apparent, from about 1,000 publications in 2013-2015 to almost 30,000 in 2018-2020, having a high open access rate of about 89.5%.

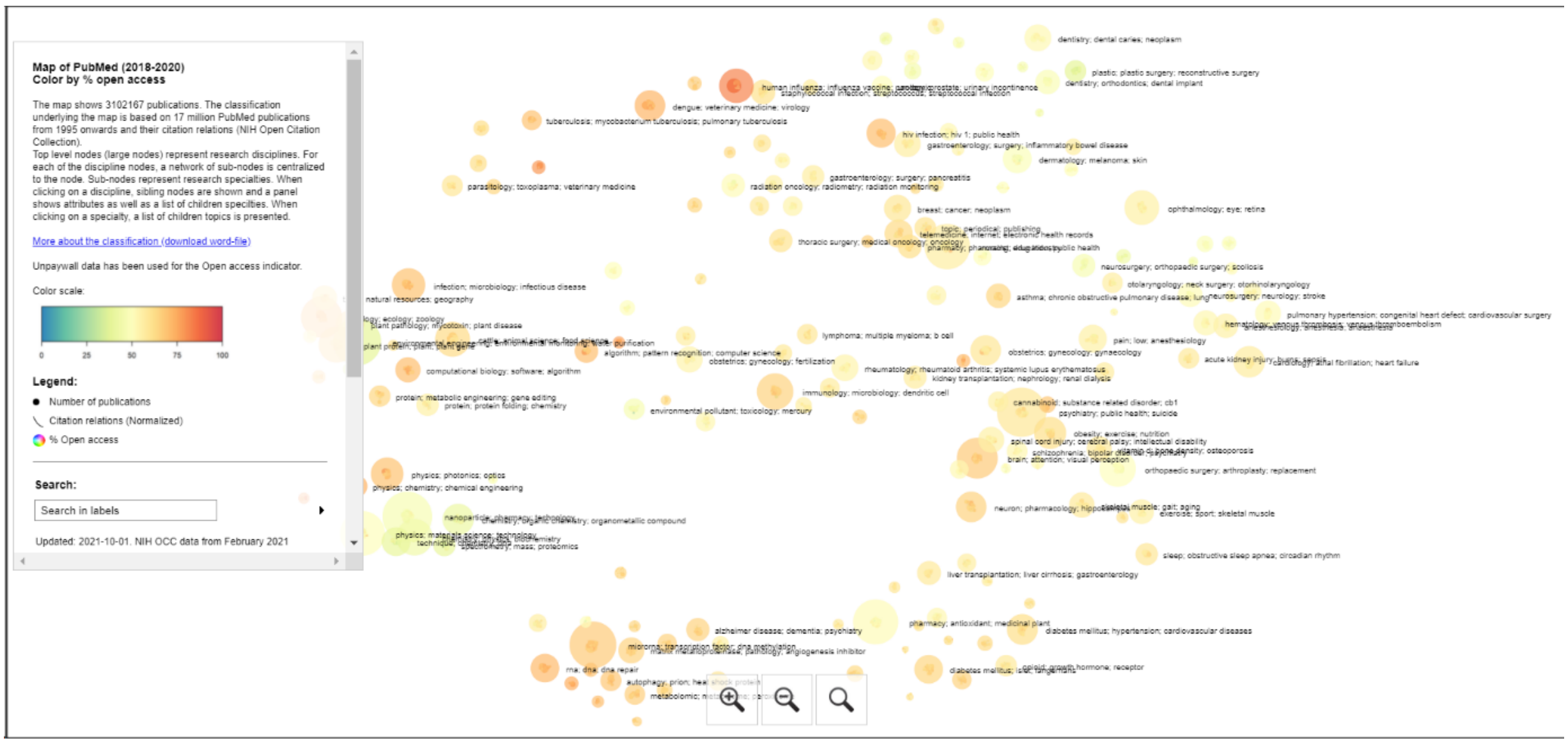

*Figure 7: Open access rates 2018-2020.*

Zooming into the 2018-2020 map reveals substantial differences between specialties. Differences can also be explored at the topic level. For instance, in immunology ("immunology; microbiology; dendritic cell") the area labeled "allele; tissue antigen; hla", concerning the role of the human leukocyte antigen (HLA) complex in the immune system, had an open access rate of only 31.8%. When we click on this node, we find a major topic engaged in gene sequencing in the HLA complex with an open access rate of only 0.6%. Following the link to PubMed and scrutinizing the list of publications reveals a focus to one journal, named "HLA", which is not an open access journal and has a low level of hybrid open access publications. This is an example of how zooming in on a particular research field can provide a more detailed understanding of the data.

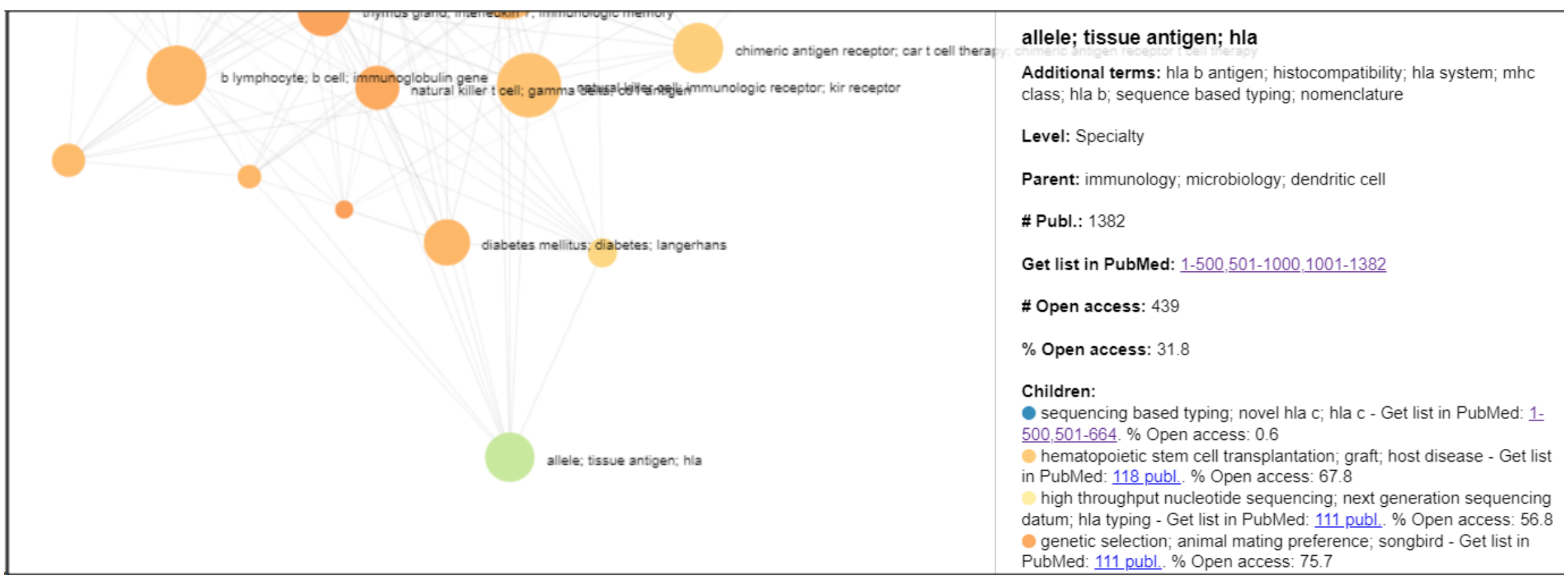

*Figure 8: Open access rates in the specialty labeled "allele; tissue antigen; hla".*

## 4.3. Coronavirus

To create a map of research related to the coronavirus pandemic that started in late 2019, I used the search query in Table 2. The query has been designed by the library at Karolinska institutet to get publications both about the disease (Covid-19) and the virus causing the disease (SARS-CoV-2).

*Table 2: Search query for Covid-19/SARS-CoV2 research*

| Covid*[tw] OR nCov[tw] OR 2019 ncov[tw] OR novel coronavirus[tw] OR novel corona virus[tw] OR " Covid-19"[All Fields] OR "Covid-2019"[All Fields] OR "severe acute respiratory syndrome coronavirus 2"[Supplementary Concept] OR "severe acute respiratory syndrome coronavirus 2"[All Fields] OR "2019-nCoV"[All Fields] OR "SARS-CoV-2"[All Fields] OR "2019nCoV"[All Fields] OR (("Wuhan"[All Fields] AND ("coronavirus"[MeSH Terms] OR "corona virus"[All Fields] OR "coronavirus"[All Fields])) AND (2019/12[PDAT] OR 2020[PDAT] OR 2021[PDAT])) |
| --- |

The search query resulted in 60953 articles from 2019 until February 2021. The Covid-19/SARS-CoV-2 map (Figure 9)[14] shows that most research related to the pandemic fuses into one specialty ("Covid 19; coronavirus infections; viral pneumonia") focused on coronaviruses, also including SARS and MERS and more general pandemic research. Nonetheless, almost half of the publications retrieved from the search query are distributed over a wide range of other specialties, including specialties in psychiatry and public health ("psychiatry; public health; suicide"), nursing and other health profession related research ("nursing; education; public health"), remote health care ("telecare; remote consultation; american telemedicine") and immunology ("immunology; microbiology; dendritic cell").

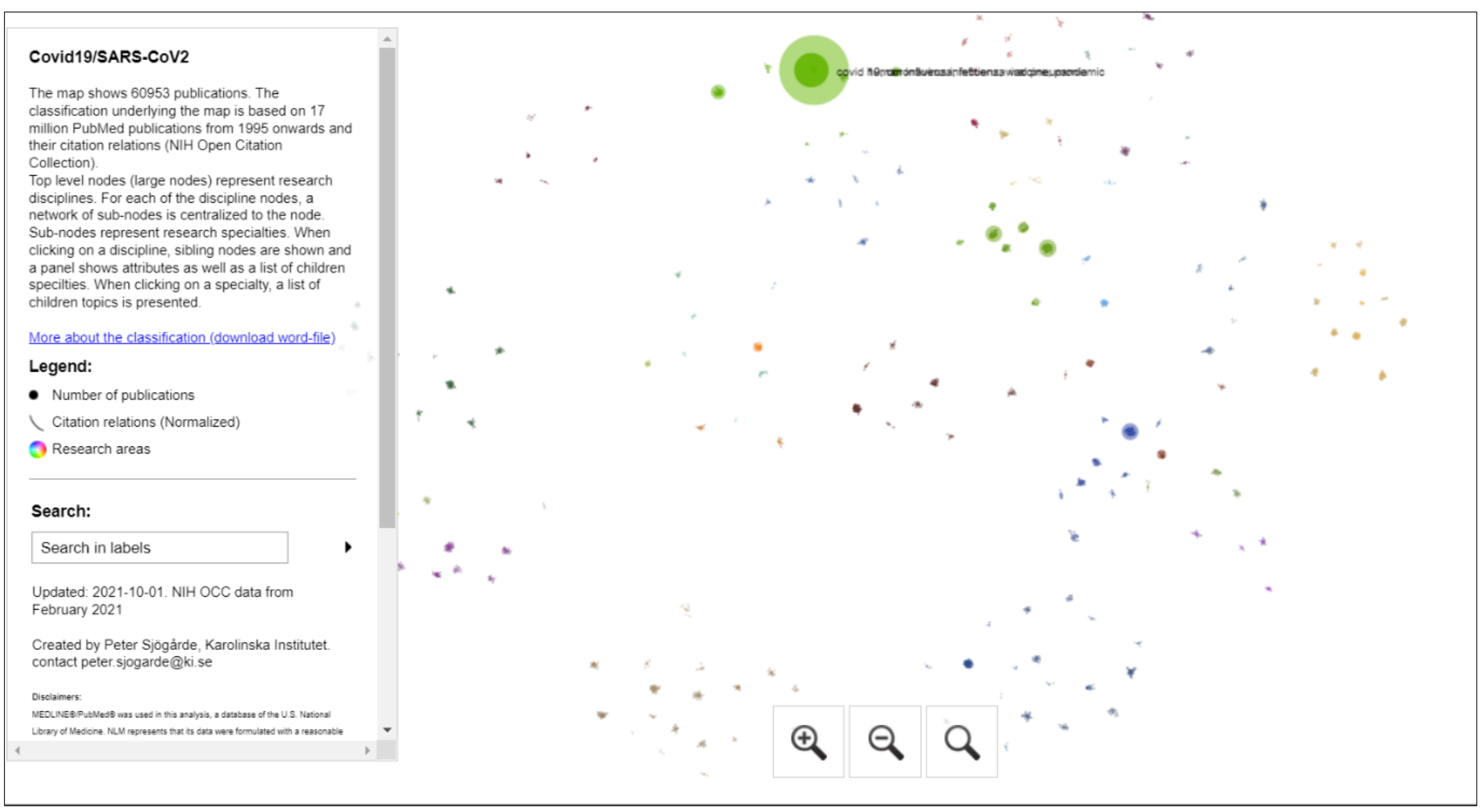

*Figure 9: Covid-19/SARS-CoV-2 research.*

Zooming into the coronavirus node (Figure 10) reveals some major topics addressing; mathematical models of the growth and spread of the disease ("basic reproduction number; serial interval; mathematical model"), effects on the cardiovascular system ("myocardial injury;

---

[14] https://petersjogarde.github.io/papers/hiervis/covid/pubs/index.html

troponin i; cardiac injury”), psychological impact of the pandemic and the measures taken to control the spread (“anxiety; depression; psychological impact”), testing (“serologic test; immunoassay; igm”), and neurological symptoms and effects (“olfaction disorders; nervous system disease; smell”).

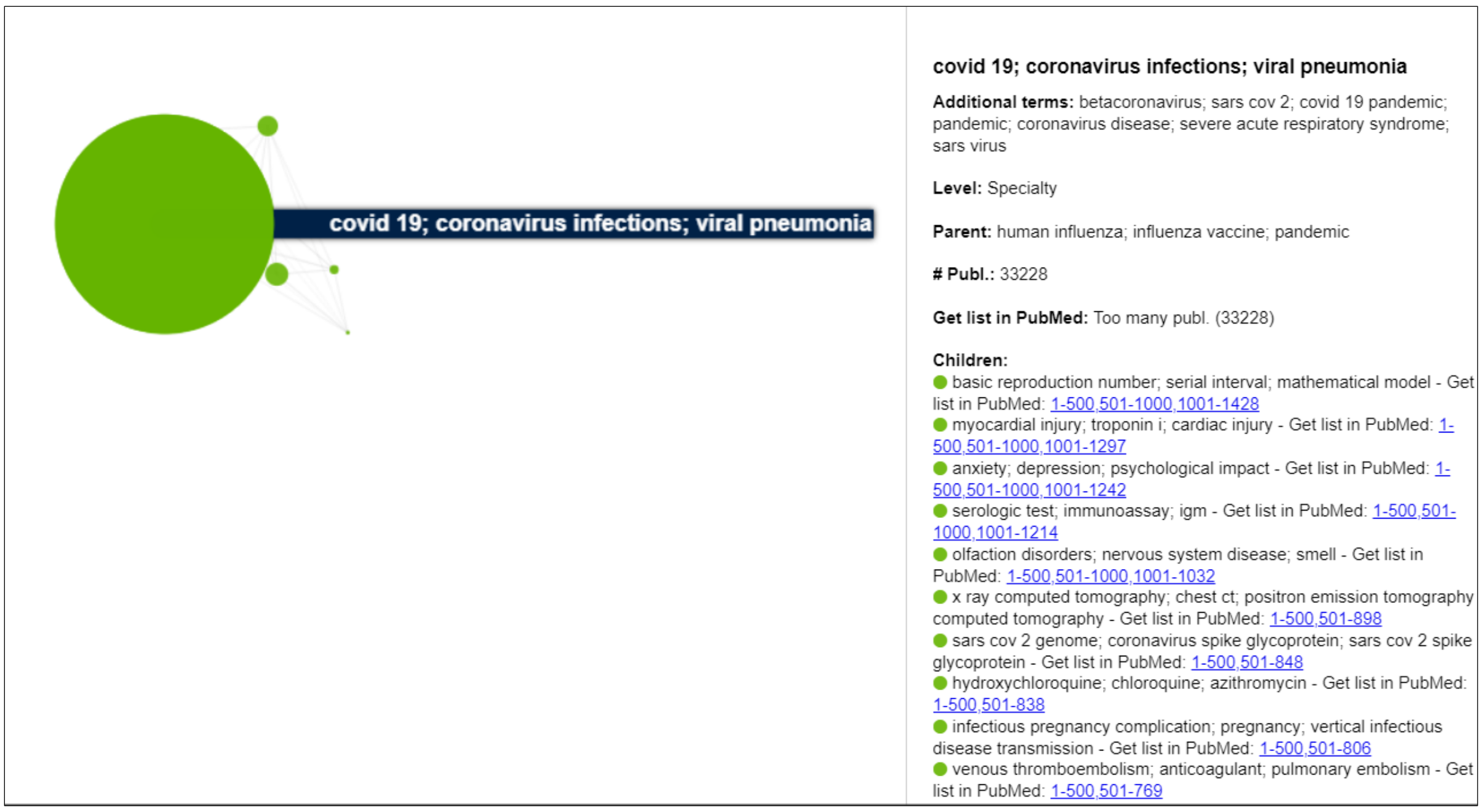

*Figure 10: Topics in coronavirus cluster (“covid 19; coronavirus infections; viral pneumonia”).*

By creating a map of research cited by the coronavirus research (Figure 11)[15], we get a picture of the research upon which the coronavirus research has been built. Node sizes are relative to the total number of publications cited by the set of coronavirus research publications and color have been set by the clusters’ average number of citations from this set. The map shows that coronavirus research is based on a wide range of areas, for example: (1) knowledge from previous coronavirus epidemics (topics addressing SARS and MERS in the specialty “covid 19; coronavirus infections; viral pneumonia”), (2) research on RNA polymerase and genetic transcription (the topic “rna dependent rna polymerase; methyltransferase; genetic transcription” in “porcine; reproductive syndrome virus; coronavirus infections”), (3) research on the angiotensin receptor used by the virus to enter the lungs (“angiotensin; angiotensin receptor; angiotensin i” in “hypertension; vascular endothelium; blood pressure”), (4) immune response prediction models (the topic “sars cov 2; immunoinformatic approach; designing” in the specialty “peptide; bacteriophage m13; b lymphocyte epitope”), and (5) prediction of how molecules bind to receptors (the topic “function; molecular docking simulation; protein ligand docking” in the specialty “chemical information; quantitative structure activity relationship; modeling”).

[15] https://petersjogarde.github.io/papers/hiervis/covid/cited/index.html

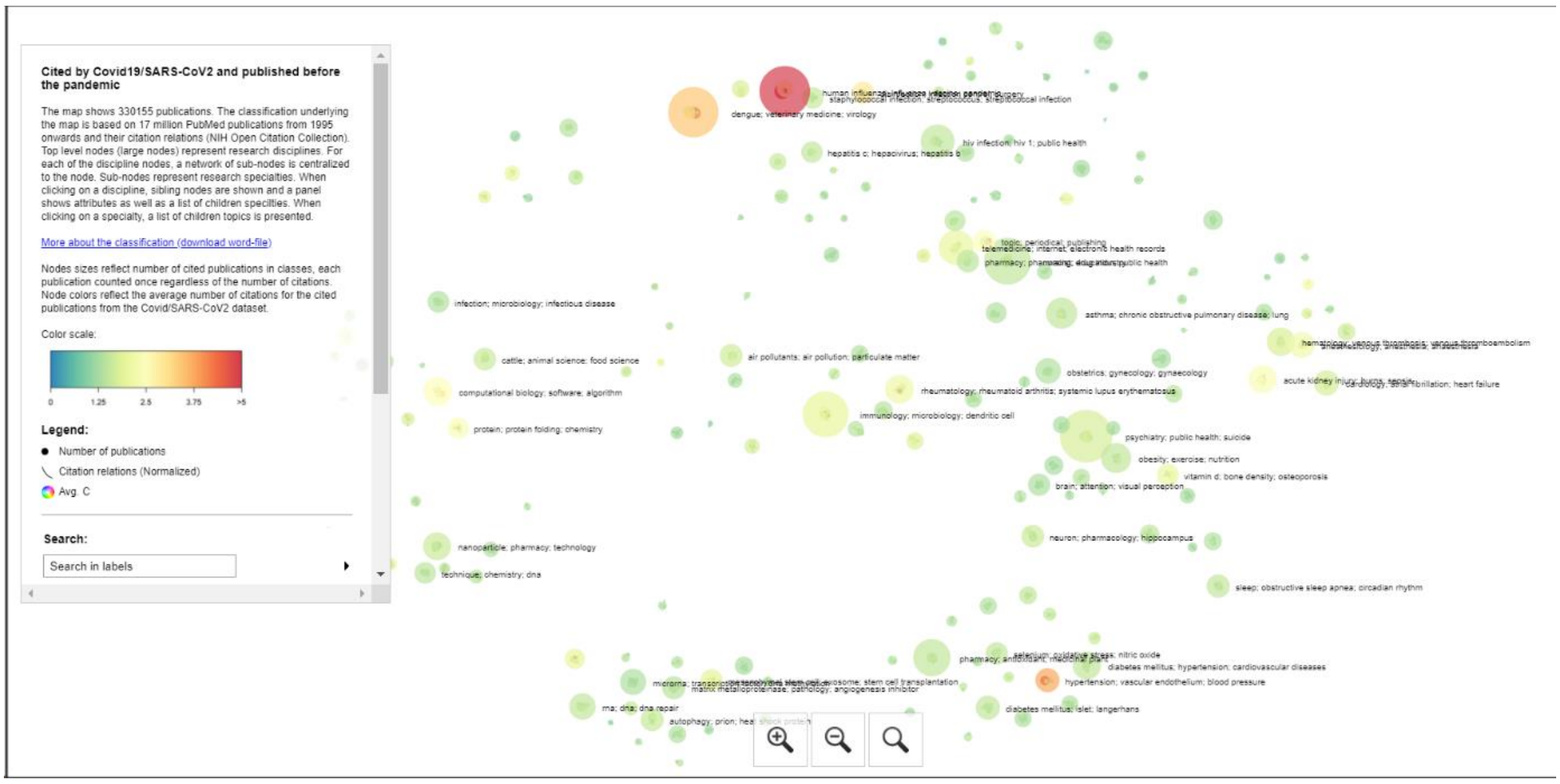

*Figure 11: Publications cited by Covid-19/SARS-CoV-2 research, published before the pandemic. Node sizes reflect the number of cited publications and node colors the average number of citations per cited publication.*

## 5 Discussion

I have shown how a publication-level classification, including both coarse and granular levels, can be used to create overlay maps of science that provide both overview and detail. To exemplify the use of such maps I have demonstrated potential utilization by revealing field differences in open access publishing and exploring the topical structure of coronavirus/Covid-19 research.

The visualizations created by the methodology put forward in this paper enables navigation of millions of articles, from broad levels down to individual articles. No existing software supports such navigation. It makes it possible to study the position of any data set in the biomedical research landscape, for example the publication output of an organization or, a journal or the publication output within a particular research field. The contents of the disciplines and specialties displayed in the visualization can be explored down to narrow topics and individual articles. Thereby, analysts and other users can get a deeper and richer understanding of the data displayed in overlay maps. Potentially, the visualization technique can be used in other applications, for example to display local maps or to visualize search results in information retrieval systems.

There are some limitations of the visualization methodology. Some of these limitations relate to the creation of the classification, some to the visualization methodology itself. Several researchers have acknowledged that different choices of methods, parameter values, relational measures and clustering algorithms results in diverse representations of science, sometimes equally valid (for examples, see Gläser et al., 2017; Waltman et al., 2020). My choice of citation relation, clustering algorithm, parameter values and labelling approach have been guided by both empirical support and practical considerations. My approach based on direct citations has the advantages to be efficient (having fewer relations than bibliographic coupling and co-citations) and to include relations to/from a large proportion of the publications in the data source, given that a comprehensive data source and a large time span are used. The direct citation approach has performed well in quantitative evaluations using large corpuses (Boyack & Klavans, 2020; Klavans & Boyack, 2017). Also the resolution parameter values have been guided by previous research (Sjögårde & Ahlgren, 2018, 2020). Nonetheless, other representations may be equally valid and express other aspects of the research landscape. For example, co-citations may be a better choice if one wants to examine the historical development of a research field, and bibliographic coupling might be preferable to display related publications in an information retrieval setting.

Labeling of the obtained clusters is a challenging task. Even though the used methodology works reasonably well, the subject orientation of clusters is sometimes hard to interpret using the cluster labels. Occasionally, other information needs to be considered by a user to understand the subject orientation of a cluster and to distinguish it from other clusters, for example additional key terms, sibling cluster labels, child cluster labels and consulting publication records in PubMed. Providing interactivity facilitates such interpretation. However, it remains unclear to what extent interpretation is a problem for users. Further work evaluating the interpretability of classifications from a user perspective is therefore needed.

The visualizations that I have presented include clusters visualized as nodes at two levels. It is possible to visualize more levels, but with the risk to make the visualization more cluttered and harder to interpret. Functionality hiding nodes at granular levels when zooming out and showing nodes when zooming in might be an option to be able to include more levels in the visualization. However, including nodes at additional levels does not necessarily help users to read and interpret the visualization. Users might for example prefer reading lists at more granular levels. Therefore, user studies are needed to develop user friendly features and to make interactive overlay maps of science easier to interpret.

The intention of this study has not been to evaluate normalization methods or layout algorithms. There might be better options to create layouts, in particular at the discipline level, both regarding normalization of citation relations and layout algorithm.

I have emphasized the clusters by multiplying relations within disciplines by a constant factor (set to 3). This procedure puts disciplines in the same cluster in proximity and improves readability. However, it may distort relations outside the cluster. The choice of the ForceAtlas layout algorithm was guided by my experience visualizing a wide range of bibliometric networks (e.g., co-author networks, MeSH-networks, co-authoring organization networks and article citation networks) in bibliometric practice at Swedish universities. ForceAtlas is implemented in the visualization software Gephi,[16] which makes it possible to try out different parameter values to facilitate the readability of a particular network and more generally to learn how to use parameter values for different kinds of networks. In my experience, the ForceAtlas algorithm creates visualizations that are interpretable and make sense, and I have received positive feedback from users on visualizations created using this layout. Alternatives to ForceAtlas are for example the OpenOrd layout algorithm used by Boyack et al. (2020), the VOS layout algorithm which is implemented in the VOSviewer software (Eck et al., 2010; Eck & Waltman, 2010), the Fruchterman-Reingold layout algorithm (Fruchterman & Reingold, 1991) and the Kamada-Kawai layout algorithm (Kamada & Kawai, 1989).

The maps created in this study have been restricted to the biomedical sciences. During the last years the amount and proportion of available bibliographic metadata has increased substantially. Future work may be extended to other research fields.

There are several technical issues related to the used visualization tool. For example, the current version of the used visualization package does not support smartphones and tablets, identification of areas of interest to a user could be facilitated by filters and improved search features, loading the visualization files is rather slow, and hyperlinks are provided in batches if a cluster includes more than 500 publications. The intention has not been to provide a perfect visualization tool, but rather to show how interactive visualizations of hierarchical classifications can provide users with enriched possibilities to explore the scientific literature. I have demonstrated that it is possible to provide maps of science that can give the user an overview of millions of publications and details down to individual publications. Such maps may constitute a valuable tool for researchers studying science, improve transparency to cluster-based citation normalization, support research management and policy making and constitute a tool for researchers to explore research of relevance to them.

[16] https://gephi.org/

## Data availability

All maps and underlying data and configuration files are available online:

**Classification and labels:**

https://doi.org/10.6084/m9.figshare.c.5610971.v1

**Base map:**

https://petersjogarde.github.io/pm_classification/2021/research_areas/index.html

**Base map files:**

https://github.com/petersjogarde/petersjogarde.github.io/tree/main/pm_classification/2021/research_areas

**Open access maps:**

2008-2010: https://petersjogarde.github.io/papers/hiervis/oa/2008-2010/index.html

2013-2015: https://petersjogarde.github.io/papers/hiervis/oa/2013-2015/index.html

2018-2020: https://petersjogarde.github.io/papers/hiervis/oa/2018-2020/index.html

**Open access files:**

https://github.com/petersjogarde/petersjogarde.github.io/tree/main/papers/hiervis/oa

**Covid-19/SARS-CoV-2 maps:**

Publications: https://petersjogarde.github.io/papers/hiervis/covid/pubs/index.html

Cited publications: https://petersjogarde.github.io/papers/hiervis/covid/cited/index.html

**Covid-19/SARS-CoV-2 files:**

https://github.com/petersjogarde/petersjogarde.github.io/tree/main/papers/hiervis/covid

## Competing interests

The author declares no competing interests.

## Funding information

Peter Sjögårde was funded by The Foundation for Promotion and Development of Research at Karolinska Institutet.

## References

Ahlgren, P., Chen, Y., Colliander, C., & van Eck, N. J. (2020). Enhancing direct citations: A comparison of relatedness measures for community detection in a large set of PubMed publications. *Quantitative Science Studies*, *1*(2), 1–17. https://doi.org/10.1162/qss_a_00027

Bassecoulard, E., & Zitt, M. (1999). Indicators in a research institute: A multi-level classification of scientific journals. *Scientometrics*, *44*(3), 323–345. https://doi.org/10.1007/BF02458483

Börner, K., Chen, C., & Boyack, K. W. (2005). Visualizing knowledge domains. *Annual Review of Information Science and Technology*, *37*(1), 179–255. https://doi.org/10.1002/aris.1440370106

Boyack, K. W. (2017). Investigating the effect of global data on topic detection. *Scientometrics*, *111*(2), 999–1015. https://doi.org/10.1007/s11192-017-2297-y

Boyack, K. W., & Klavans, R. (2010). Co-citation analysis, bibliographic coupling, and direct citation: Which citation approach represents the research front most accurately? *Journal of the American Society for Information Science and Technology*, *61*(12), 2389–2404. https://doi.org/10.1002/asi.21419

Boyack, K. W., & Klavans, R. (2020). A comparison of large-scale science models based on textual, direct citation and hybrid relatedness. *Quantitative Science Studies*, 1–16. https://doi.org/10.1162/qss_a_00085

Boyack, K. W., & Klavans, R. (2018). Accurately identifying topics using text: Mapping PubMed. *STI 2018 Conference Proceedings*, *23*. https://openaccess.leidenuniv.nl/handle/1887/65319

Boyack, K. W., Klavans, R., & Börner, K. (2005). Mapping the backbone of science. *Scientometrics*, *64*(3), 351–374. https://doi.org/10.1007/s11192-005-0255-6

Boyack, K. W., Newman, D., Duhon, R. J., Klavans, R., Patek, M., Biberstine, J. R., Schijvenaars, B., Skupin, A., Ma, N., & Börner, K. (2011). Clustering more than two million biomedical publications: Comparing the accuracies of nine text-based similarity approaches. *PLOS ONE*, *6*(3), e18029. https://doi.org/10.1371/journal.pone.0018029

Boyack, K. W., Smith, C., & Klavans, R. (2020). A detailed open access model of the PubMed literature. *Scientific Data*, *7*(1), 408. https://doi.org/10.1038/s41597-020-00749-y

Donner, P. (2021). Validation of the Astro dataset clustering solutions with external data. *Scientometrics*, *126*(2), 1619–1645. https://doi.org/10.1007/s11192-020-03780-3

Eck, N. J. van, & Waltman, L. (2010). Software survey: VOSviewer, a computer program for bibliometric mapping. *Scientometrics*, *84*(2), 523–538. https://doi.org/10.1007/s11192-009-0146-3

Eck, N. J. van, Waltman, L., Dekker, R., & Berg, J. van den. (2010). A comparison of two techniques for bibliometric mapping: Multidimensional scaling and VOS. *Journal of the American Society for Information Science and Technology*, *61*(12), 2405–2416. https://doi.org/10.1002/asi.21421

Fruchterman, T. M. J., & Reingold, E. M. (1991). Graph drawing by force-directed placement. *Software: Practice and Experience*, *21*(11), 1129–1164. https://doi.org/10.1002/spe.4380211102

Glänzel, W., & Schubert, A. (2003). A new classification scheme of science fields and subfields designed for scientometric evaluation purposes. *Scientometrics*, *56*(3), 357–367. https://doi.org/10.1023/A:1022378804087

Gläser, J., Glänzel, W., & Scharnhorst, A. (2017). Same data—different results? Towards a comparative approach to the identification of thematic structures in science. *Scientometrics*, *111*(2), 981–998. https://doi.org/10.1007/s11192-017-2296-z

Haunschild, R., Schier, H., Marx, W., & Bornmann, L. (2018). Algorithmically generated subject categories based on citation relations: An empirical micro study using papers on overall water splitting. *Journal of Informetrics*, *12*(2), 436–447. https://doi.org/10.1016/j.joi.2018.03.004

Held, M., Laudel, G., & Gläser, J. (2021). Challenges to the validity of topic reconstruction. *Scientometrics*. https://doi.org/10.1007/s11192-021-03920-3

Hutchins, B. I., Baker, K. L., Davis, M. T., Diwersy, M. A., Haque, E., Harriman, R. M., Hoppe, T. A., Leicht, S. A., Meyer, P., & Santangelo, G. M. (2019). The NIH Open Citation Collection: A public access, broad coverage resource. *PLOS Biology*, *17*(10), e3000385. https://doi.org/10.1371/journal.pbio.3000385

iCite, Hutchins, B. I., & Santangelo, G. (2019). *ICite Database Snapshots (NIH Open Citation Collection)*. The NIH Figshare Archive. Collection. https://doi.org/10.35092/yhjc.c.4586573

Jacomy, M., Venturini, T., Heymann, S., & Bastian, M. (2014). ForceAtlas2, a continuous graph layout algorithm for handy network visualization designed for the Gephi software. *PLoS ONE*, *9*(6), e98679. https://doi.org/10.1371/journal.pone.0098679

Kamada, T., & Kawai, S. (1989). An algorithm for drawing general undirected graphs. *Information Processing Letters*, *31*(1), 7–15. https://doi.org/10.1016/0020-0190(89)90102-6

Kay, L., Newman, N., Youtie, J., Porter, A. L., & Rafols, I. (2014). Patent overlay mapping: Visualizing technological distance. *Journal of the Association for Information Science and Technology*, *65*(12), 2432–2443. https://doi.org/10.1002/asi.23146

Klaine, S. J., Koelmans, A. A., Horne, N., Carley, S., Handy, R. D., Kapustka, L., Nowack, B., & von der Kammer, F. (2017). Paradigms to assess the environmental impact of manufactured nanomaterials. *Integrated Environmental Assessment and Management*, 3–14. https://doi.org/10.1002/etc.733@10.1002/(ISSN)1551-3793.InRecognitionofaDistinguishedCareerSteveKlaine

Klavans, R., & Boyack, K. W. (2011). Using global mapping to create more accurate document-level maps of research fields. *Journal of the American Society for Information Science and Technology*, *62*(1), 1–18. https://doi.org/10.1002/asi.21444

Klavans, R., & Boyack, K. W. (2017). Which type of citation analysis generates the most accurate taxonomy of scientific and technical knowledge? *Journal of the Association for Information Science and Technology*, *68*(4), 984–998. https://doi.org/10.1002/asi.23734

Leydesdorff, L. (2004). Clusters and maps of science journals based on bi-connected graphs in the journal citation reports. *J. Documentation*. https://doi.org/10.1108/00220410410548144

Leydesdorff, L. (2006). Can scientific journals be classified in terms of aggregated journal-journal citation relations using the Journal Citation Reports? *Journal of the American Society for Information Science and Technology*, *57*(5), 601–613. https://doi.org/10.1002/asi.20322

Leydesdorff, L., Moya-Anegón, F. de, & Guerrero-Bote, V. P. (2015). Journal maps, interactive overlays, and the measurement of interdisciplinarity on the basis of Scopus data (1996–2012). *Journal of the Association for Information Science and Technology*, *66*(5), 1001–1016. https://doi.org/10.1002/asi.23243

Mai, J. (2011). The modernity of classification. *Journal of Documentation*, *67*(4), 710–730. https://doi.org/10.1108/00220411111145061

Manning, C. D., Surdeanu, M., Bauer, J., Finkel, J. R., Bethard, S., & McClosky, D. (2014). The Stanford CoreNLP natural language processing toolkit. *Proceedings of 52nd Annual Meeting of the Association for Computational Linguistics: System Demonstrations*, *52*, 55–60. https://doi.org/10.3115/v1/P14-5010

Moya-Anegón, F., Vargas-Quesada, B., Herrero-Solana, V., Chinchilla-Rodríguez, Z., Corera-Álvarez, E., & Munoz-Fernández, F. J. (2004). A new technique for building maps of large scientific domains based on the cocitation of classes and categories. *Scientometrics*, *61*(1), 129–145. https://doi.org/10.1023/B:SCIE.0000037368.31217.34

Petrovich, E. (2020). Science mapping. In *Encyclopedia of Knowledge Organization*. https://www.isko.org/cyclo/science_mapping

Rafols, I., Porter, A. L., & Leydesdorff, L. (2010). Science overlay maps: A new tool for research policy and library management. *Journal of the American Society for Information Science and Technology*, *61*(9), 1871–1887. https://doi.org/10.1002/asi.21368

RoRI Institute, Waltman, L., Rafols, I., van Eck, N. J., & Yegros Yegros, A. (2019). *Supporting priority setting in science using research funding landscapes* (Report No. 1; RoRI Working Paper). Research on Research Institute. https://doi.org/10.6084/m9.figshare.9917825.v1

Rotolo, D., Rafols, I., Hopkins, M. M., & Leydesdorff, L. (2017). Strategic intelligence on emerging technologies: Scientometric overlay mapping. *Journal of the Association for Information Science and Technology*, *68*(1), 214–233. https://doi.org/10.1002/asi.23631

Sjögårde, P., & Ahlgren, P. (2018). Granularity of algorithmically constructed publication-level classifications of research publications: Identification of topics. *Journal of Informetrics*, *12*(1), 133–152. https://doi.org/10.1016/j.joi.2017.12.006

Sjögårde, P., & Ahlgren, P. (2020). Granularity of algorithmically constructed publication-level classifications of research publications: Identification of specialties. *Quantitative Science Studies*, *1*(1), 207–238. https://doi.org/10.1162/qss_a_00004

Sjögårde, P., Ahlgren, P., & Waltman, L. (2021). Algorithmic labeling in hierarchical classifications of publications: Evaluation of bibliographic fields and term weighting approaches. *Journal of the Association for Information Science and Technology*, *72*(7), 853–869. https://doi.org/10.1002/asi.24452

Smiraglia, R. P., & van den Heuvel, C. (2013). Classifications and concepts: Towards an elementary theory of knowledge interaction. *Journal of Documentation; Bradford*, *69*(3), 360–383. http://dx.doi.org.ezproxy.its.uu.se/10.1108/JD-07-2012-0092

Šubelj, L., van Eck, N. J., & Waltman, L. (2016). Clustering scientific publications based on citation relations: A systematic comparison of different methods. *PLOS ONE*, *11*(4), e0154404. https://doi.org/10.1371/journal.pone.0154404

Tang, L., & Shapira, P. (2011). China–US scientific collaboration in nanotechnology: Patterns and dynamics. *Scientometrics*, *88*(1), 1. https://doi.org/10.1007/s11192-011-0376-z

Toutanova, K., Klein, D., Manning, C. D., & Singer, Y. (2003). Feature-rich part-of-speech tagging with a cyclic dependency network. *Proceedings of the 2003 Conference of the North American Chapter of the Association for Computational Linguistics on Human Language Technology*, *1*, 173–180. https://doi.org/10.3115/1073445.1073478

Toutanova, K., & Manning, C. D. (2000). Enriching the knowledge sources used in a maximum entropy part-of-speech tagger. *TEST Proceedings of the 2000 Joint SIGDAT Conference on Empirical Methods in Natural Language Processing and Very Large Corpora: Held in Conjunction with the 38th Annual Meeting of the Association for Computational Linguistics*, *13*, 63–70. https://doi.org/10.3115/1117794.1117802

Traag, V. A., Waltman, L., & Eck, N. J. van. (2019). From Louvain to Leiden: Guaranteeing well-connected communities. *Scientific Reports*, *9*(1), 5233. https://doi.org/10.1038/s41598-019-41695-z

van Eck, N. J., & Waltman, L. (2014). Visualizing Bibliometric Networks. In Y. Ding, R. Rousseau, & D. Wolfram (Eds.), *Measuring Scholarly Impact: Methods and Practice* (pp. 285–320). Springer International Publishing. https://doi.org/10.1007/978-3-319-10377-8_13

van Eck, N. J., Waltman, L., Noyons, E. C. M., & Buter, R. K. (2010). Automatic term identification for bibliometric mapping. *Scientometrics*, *82*(3), 581–596. https://doi.org/10.1007/s11192-010-0173-0

Velden, T., Boyack, K. W., Gläser, J., Koopman, R., Scharnhorst, A., & Wang, S. (2017). Comparison of topic extraction approaches and their results. *Scientometrics*, *111*(2), 1169–1221. https://doi.org/10.1007/s11192-017-2306-1

Waltman, L., Boyack, K. W., Colavizza, G., & van Eck, N. J. (2020). A principled methodology for comparing relatedness measures for clustering publications. *Quantitative Science Studies*, *1*(2), 691–713. https://doi.org/10.1162/qss_a_00035

Waltman, L., & van Eck, N. J. (2012). A new methodology for constructing a publication-level classification system of science. *Journal of the American Society for Information Science and Technology*, *63*(12), 2378–2392. https://doi.org/10.1002/asi.22748

Zitt, M., Lelu, A., Cadot, M., & Cabanac, G. (2019). Bibliometric delineation of scientific fields. *Springer Handbook of Science and Technology Indicators*. https://doi.org/10.1007/978-3-030-02511-3_2